# EPR studies of manganese centers in SrTiO$_3$: Non-Kramers Mn$^{3+}$ ions and spin-spin coupled Mn$^{4+}$ dimers

**D V Azamat[1], A Dejneka[1], J Lancok[1], V A Trepakov[1, 2], L Jastrabik[1] and A G Badalyan[2]**

[1] Institute of Physics AS CR, 182 21, Prague 8, Czech Republic

[2] Ioffe Physical-Technical Institute, Russian Academy of Sciences, 194021, St. Petersburg, Russia

E-mail: azamat@fzu.cz

**Abstract**

X- and Q-band electron paramagnetic resonance (EPR) study is reported on the SrTiO$_3$ single crystals doped with 0.5-at.% MnO. EPR spectra originating from the $S = 2$ ground state of Mn$^{3+}$ ions are shown to belong to the three distinct types of Jahn-Teller centres. The ordering of the oxygen vacancies due to the reduction treatment of the samples and consequent formation of oxygen vacancy associated Mn$^{3+}$ centres are explained in terms of the localized charge compensation. The EPR spectra of SrTiO$_3$: Mn crystals show the presence of next nearest neighbor exchange coupled Mn$^{4+}$ pairs in the <110> directions.

## 1. Introduction

Transition metal doped perovskites has attracted increasing interest due to a variety of physical phenomena such as superconductivity, colossal magnetoresistance and metal-to-isolator transition [1-4]. Strontium titanate, the so called incipient ferroelectric with perovskite structure, has become in the focus of interest in attempt to combine magnetic and ferroelectric properties of the material [5-11]. In this work we use EPR spectroscopy for the accurate determination of different oxidation states and local environment of manganese impurities in $SrTiO_3$. Trivalent manganese ions in $SrTiO_3$ crystals look like a model ones to study exchange-coupled systems in mixed-valent manganese oxides [12-17]. EPR spectra of $Mn^{3+}$ complexes in oxide crystals have been poorly [18, 19] investigated than those of $3d^4$ non-Kramers iron ($Fe^{4+}$ ions) [20-22]. These ions exhibit large zero field splittings due to the influence of the Jahn-Teller effect on the $^5E_g$ ground state in octahedral symmetry. An EPR centre attributed to the $Mn^{3+}$-$V_O$ complex was reported by Serway et al [19], only the transition within $M_S = 2 \leftrightarrow M_S = -2$ states of electron spin $S = 2$ system was observed. A strong axial field is created by a nearest-neighbour oxygen vacancy $V_O$. So far, the lack of information on the $Mn^{3+}$ centres in $SrTiO_3$ has precluded the detailed analysis for that system. Owing to the large interest in the characterisation of $Mn^{3+}$ complexes in perovskites, we decided to perform EPR study on this system.

In this paper we report X- and Q-band EPR measurements on $SrTiO_3$: Mn single crystals. The EPR spectrum of $SrTiO_3$ crystals doped with manganese contains resonances from cubic $Mn^{4+}$ and $Mn^{2+}$ ions as well as from several types of $Mn^{3+}$ centres. The accurate values of the 2-nd and 4-th rank crystal field tensor components have been obtained for distinct types of $Mn^{3+}$ centres. The peculiar features of $Mn^{3+}$ centres were established by the studies of hyperfine interaction with 100% abundant $^{55}Mn$ nuclei (nuclear spin of $I = 5/2$). The spectra of distinct types of $Mn^{3+}$ centres differ greatly in the magnitude of their $^{55}Mn$ hyperfine fields.

The careful analysis of spin-spin coupling effects may provide some useful physical information on structural and electronic properties of $SrTiO_3$. When crystals of $SrTiO_3$ are heavily doped with fourvalent manganese there is a tendency for the $Mn^{4+}$ ions to form exchange coupled pairs. The EPR spectra indicate the formation of $Mn^{4+}$ dimer centers lying along the <110> directions (in the next nearest neighbor position). The coupling in the ground state is expected to be ferromagnetic.

## 2. Experimental procedure

The single crystals of $SrTiO_3$ were grown in Furuuchi Chemical Corporation by use of Verneuil technique through the melting of $SrTiO_3$ and 0.5% wt of MnO crystalline powders. Some of the samples were annealed for 6 h at 1000 C in $Ar/H_2$ (5%) reduction atmosphere. EPR measurements were performed with a Bruker ELEXSYS 580 spectrometer at X- and Q-band frequencies in the temperature range 65-300K.

## 3. Results and Discussion

### 3.1. Non-Kramers $Mn^{3+}$ ions

In addition to the cubic $Mn^{4+}$ ($3d^3$) and $Mn^{2+}$ ($3d^5$) spectra, that are already known, the spectra of strong axially distorted $Mn^{3+}$ ($3d^4$) ions have been found (see Fig. 1). These manganese centers as well as a variety of the other $3d$ impurities in $SrTiO_3$ are typically substituted for $Ti^{4+}$ sites in the octahedral coordination [11, 19, 23-25].

**Table 1.** The Spin Hamiltonian parameters of $Mn^{3+}$ centres in $SrTiO_3$ single crystals (in $10^{-4}$ cm$^{-1}$, 295K).

| centre | $b_2^0$ | $b_4^0$ | $b_4^4$ | $g_\parallel$ | $g_\perp$ | $A_\parallel$ | $A_\perp$ |
|---|---|---|---|---|---|---|---|
| $Mn^{3+}$-$V_O$ (I) | 28440.0 | 170.0 | -3800.0 | 1.986 | 1.997 | -38.0 | -64.0 |
| $Mn^{3+}$-X | 27550.0 | 170.0 | -1350.0 | 1.986 | 1.997 | -10.4 | -29.8 |
| $Mn^{3+}$-$V_O$ (II) | 26340.0 | 170.0 | -1350.0 | 1.986 | 1.997 | -28.0 | -45.0 |

Practically, the studies of the angular dependencies of EPR spectra reveal three types of the $Mn^{3+}$ centers with the spin $S = 2$. These centers has axial symmetry with the main axis along to the [100]-type directions.

Figure 2 shows the axial $Mn^{3+}$ centres for applied magnetic field direction B||[100]. The $Mn^{3+}$ spectra consists of 2 ↔ -2 and 1 ↔ -1 fine structure transitions within $^5B_1$ ground state. We use the following spin Hamiltonian with symmetry $D_{4h}$ to describe the ground state energy splitting of $Mn^{3+}$ centres:

$$H = g_\parallel \beta_e B_z S_z + g_\perp \beta_e (B_x S_x + B_y S_y) + B_{20} \overline{O}_0^2 + B_{40} \overline{O}_0^4 + B_{44} \overline{O}_4^4 \\ + A_\parallel S_z I_z + A_\perp (S_x I_x + S_y I_y) \quad (1)$$

$$\overline{O}_0^l = T_0^l, \quad \overline{O}_m^l = T_{-m}^l + (-1)^m T_m^l$$

where $S = 2$, the first two terms describes the Zeeman interaction, the last two ones correspond to the hyperfine (HF) interactions; $\beta_e$ is the Bohr magneton; $g$ is the g-tensor; $A$ is the hyperfine (HF) interaction tensor; $B_{20} = \frac{2}{\sqrt{6}} b_2^0$, $B_{40} = \frac{\sqrt{70}}{30} b_4^0$, $B_{44} = \frac{1}{30} b_4^4$ are the scalar fine structure constants; $T_m^l$ are spherical tensor operators [28] and $b_l^m$ are Stevens coefficients; $x, y, z$, the crystallographic axes.

Figure 3 shows the angular variation of the spectra measured in the (100) plane at the temperature T=150K. The magnetic multiplicity of EPR spectra $K_M = 3$ corresponds with three possible axial distortions along [100] type of directions. The curves on Fig. 3 have been calculated by diagonalization of the spin Hamiltonian (1) by using a program similar to that described in [28]. The characteristic EPR road map has shown just for centers designated as $Mn^{3+}$-$V_O$ (II) because the corresponding fine structure transitions of $Mn^{3+}$-$V_O$(I) and $Mn^{3+}$-X centers follows nearby through the complete angular dependence. The parameters of the spin Hamiltonian (1) for the $Mn^{3+}$-$V_O$(I), $Mn^{3+}$-X and $Mn^{3+}$-$V_O$ (II) centers are listed in Table 1. In Fig. 4 it is shown the hyperfine structure of $Mn^{3+}$ spectra are well resolved at the temperature of 295K and has narrow resonance linewidths of about 1.2 mT. The $Mn^{3+}$ hyperfine interaction is axially symmetric around a [100] direction as well as g-tensor main axis. The weaker HF interaction is for parallel magnetic-field orientation with respect to the tetragonal axis of the center (the local z axis), while the stronger HF interaction is in the equatorial plane perpendicular to the local z axis of the center.

High spin $Mn^{3+}$ free ion term is $^5D$ in octahedral environment. It gives $^5E$ ground state that is not split by spin-orbit coupling at first order of perturbation theory [26, 27]. A strong axial field results in lowering the symmetry to $D_{4h}$. This give rise to $S = 2$ high spin configuration $(t_{2g})^3(e_g)^1$ with an orbital singlet $^5B_1$ $(3z^2-r^2)$ ground state. The stabilization of ground state $3d(3z^2-r^2)$ orbital results in the shift the four oxygen ions in the equatorial plane towards the central $Mn^{3+}$ ion and the moving of two oxygen ions away making elongated octahedral. The inspection of the values of hyperfine structure constants (see Table 1) of the centers show that although the ground state $S = 2$ is the same as that for $Mn^{3+}$ the actual spin density of different types of the centers is rather delocalized among surrounding ions due to strong covalent bonding. The anisotropic hyperfine splitting for the $^5B_1$ state of different types of $Mn^{3+}$ centers can be evaluated by use of the values from Table 1. For an $3d(3z^2-r^2)$ ground state the relations are the following [26, 29-32]:

$$A_\parallel = \frac{1}{7}P - a + P\Delta g_\parallel, \qquad A_\perp = -\frac{1}{14}P - a + P\Delta g_\perp, \qquad (2a)$$

$$P = g_e g_N \beta_e \beta_N \langle r^{-3} \rangle,$$

$$\chi = \frac{4\pi}{S}\langle \psi | \Sigma_i \delta(r_i) s_{zi} | \psi \rangle = -\frac{3}{2}\left(\frac{hca_0^3}{g_e g_N \beta_e \beta_N}\right)a, \qquad (2b)$$

where isotropic contact term $\chi$ is the value of the density of unpaired spins at the nucleus; the $\parallel$ and $\perp$ symbols denote the parallel and perpendicular components of HF interaction tensor with respect to the local z axis of the center. Because the $g_\parallel$ and $A_\parallel$ parameters are smaller than $g_\perp$ and $A_\perp$ (see Table 1) the ground state $3d(3z^2-r^2)$ orbital is more likely rather than $3d(x^2-y^2)$. The adoption of negative signs of hyperfine interaction constants (see Table 1) is a reasonable choice which leads to negative contact term $\chi$ coming from the core polarization. By use of the data in Table 1 we find the values of $P$, $a$ and $\chi$ listed in the Table 2 (as evaluated by use of formulae (2)). The $P$ value for $Mn^{3+}$- X centre is appreciably smaller than for $Mn^{3+}$- $V_O$ (I, II) ones. The $Mn^{3+}$ HF interaction data in $SrTiO_3$ yields quite different magnitudes of hyperfine structure constants from that found for $Mn^{3+}$ complexes in titanium oxide ($TiO_2$) crystals [12]. The essential reduction of contact interaction $\chi$ values exhibits the strong covalent bonding of the $Mn^{3+}$ impurities. The main contribution to the contact term comes from the spin polarization of the core $s$-shells due to the exchange interaction with the $d$-electron spin states [33, 34].

Let us consider the possible microscopic models of $Mn^{3+}$ centers (see Fig. 5) based on the present experimental data. All the spectra are characterized by $g_\parallel < g_\perp < 2$ in the cubic phase. Therefore it results from $^5B_1(3z^2-r^2)$ state of $Mn^{3+}$ ion in the presence of a strong positive axial crystal field. All the $Mn^{3+}$ centers are very stable at room temperature. The dominant charge compensation process for $Mn^{3+}$ ions is the formation of the oxygen vacancies at the nearest or next nearest anionic positions along [100] type of directions.

We suppose the $Mn^{3+}$ -$V_O$ (I) center is caused by an oxygen vacancy $V_O$ in immediate neighborhood of the $Mn^{3+}$ ion where empty oxygen vacancy stabilize a $3d(3z^2-r^2)$ orbital of Jahn-Teller active defect. Thus, the EPR spectra of $Mn^{3+}$ -$V_O$ (I) ions of axial symmetry are analogous to the corresponding spectra of $Fe^{3+}$ with an oxygen vacancy in the nearest environment [35]. Other example of nearest $V_O$ charge compensated defect is $Ni^{3+}$-$V_O$ [36]. The EPR transitions within the $|\pm2\rangle$ non-Kramers doublet of $Mn^{3+}$-$V_O$ (I) pair centers with the nearest neighbor oxygen vacancy was observed for the first time by Serway et al [19].

It should be pointed out the relative concentrations of the $Mn^{3+}$-$V_O$(I) and $Mn^{3+}$-$V_O$(II) centers depend on the thermal treatment of the sample in reduced atmosphere. Figure 4 shows the EPR spectra corresponds with 1 ↔ -1 transition of as grown and additionally reduced manganese-doped $SrTiO_3$ with the applied magnetic field parallel to the [110] axis. The relative change of EPR intensity of $Mn^{3+}$-$V_O$(I) and $Mn^{3+}$-$V_O$(II) centers due to additional reduction of the samples supports to conclusion that the charge compensators like oxygen vacancies are mobile defects in respect to the $Mn^{3+}$ sites (under annealing process).

**Table 2.** Hyperfine coupling parameters of $Mn^{3+}$ centres in $SrTiO_3$ and $TiO_2$ single crystals

| centre | $10^4 P$ (cm$^{-1}$) | $10^4 a$ (cm$^{-1}$) | $\chi$ (a.u.) | ref. |
|---|---|---|---|---|
| SrTiO$_3$: Mn$^{3+}$-V$_O$ (I) | 121.3 | 55.3 | -1.88 | * |
| SrTiO$_3$: Mn$^{3+}$-X | 90.5 | 23.3 | -0.79 | * |
| SrTiO$_3$: Mn$^{3+}$- V$_O$ (II) | 79.3 | 39.3 | -1.33 | * |
| TiO$_2$: Mn$^{3+}$ | 140.0 | 73.0 | -2.45 | [18] |

\* present study

We suppose, there is no oxygen vacancy V$_O$ in the immediate neighborhood of the centers designated as Mn$^{3+}$-V$_O$(II) in Table 1. In this case, a large axial field term results from a static Jahn-Teller distortion. In the SrTiO$_3$ structure the isolated oxygen vacancy has two nearest Ti$^{4+}$ ions. The formation of neutral oxygen vacancies is accompanied by the trapping of two electrons [37]. In the case of the oxygen vacancy next nearest to Mn$^{3+}$ ion the Mn$^{3+}$- O- Ti$^{3+}$- Vo complex is formed with axial distortion along [100] direction. This model is similar to that recently presented for Cr$^{3+}$ charge compensated complex in SrTiO$_3$ [38]. Oxygen vacancy concentration could be increased by the annealing treatment of as grown sample in reducing atmosphere. Such procedure results in the relative increase of EPR signal of Mn$^{3+}$- V$_O$ (I) centers consisting of the nearest-neighbor neutrally charged oxygen vacancy (see Fig. 4). The main factors governing the distribution of trivalent Mn$^{3+}$ impurities in the SrTiO$_3$ crystals appear to be electrostatic ones.

We propose the tentative model of Mn$^{3+}$-X centers as Mn$^{3+}$ nearest to Nb$^{5+}$ ion substituted for Ti site. The Nb traces are than compensated by Mn$^{3+}$ ions in the samples. A quite weak intensity of EPR signals of Mn$^{3+}$-X centers is not visibly changed by the thermal treatment of the samples in reduced atmosphere. This is consistent with assumption that Mn$^{3+}$ ion is a probable compensator for Nb$^{5+}$ donor. Further experimental work shall concentrate on the Mn$^{3+}$-X center to confirm the model suggested by the present study.

*3.2. Exchange-coupled Mn$^{4+}$ pairs*

In as grown crystals the cubic Mn$^{4+}$ (as well as cubic Mn$^{2+}$) dominates EPR spectrum which indicates that the majority of the Mn$^{4+}$ ions enter the lattice as isolated centres. The EPR of isolated Mn$^{4+}$ ions have been described [24] by the spin Hamiltonian for a 3$d^3$ electron system ($I$=5/2) in octahedral environment:

$$H_i = g\beta \mathbf{B} \cdot \mathbf{S} + D_c\left[\mathbf{S}_z^2 - \frac{1}{3}\mathbf{S}(\mathbf{S}+1)\right] + A\mathbf{S} \cdot \mathbf{I}, \qquad (3)$$

where the axial zero-field splitting $D_c \sim 0$ at room temperature and therefore EPR spectrum is almost isotropic. The spin Hamiltonian parameters are given in Table 3. When two $Mn^{4+}$ ions with electron spin $S=3/2$ are spin-spin coupled the dimer spin states are characterized by the total spin quantum number $\mathbf{\Sigma} = \mathbf{S_1} + \mathbf{S_2} = 0, 1, 2, 3$ [26, 32, 39]. The EPR spectra of $SrTiO_3$ crystals heavily doped with $Mn^{4+}_{Ti}$ ($3d^3$) contain resonance lines which can be assigned to the total electron spin quantum number $\Sigma = 1$ of exchange-coupled pair. Figure 2 show the spectrum of $Mn^{4+}$- $Mn^{4+}$ in $SrTiO_3$ under applied magnetic field parallel to [100] direction. The isolated ion spectrum of $Mn^{4+}$ is very much more intense than the pair spectrum. Suggesting the isotropic part of exchange energy is large compared to the other magnetic interactions the X-band EPR angular dependence from the $\Sigma = 1$ state can be well described by the spin Hamiltonian [39]:

$$H = \beta \mathbf{B} \cdot \mathbf{g} \cdot \mathbf{\Sigma} + D_\Sigma \left[ \mathbf{\Sigma}_z^2 - \frac{1}{3}\Sigma(\Sigma+1) \right] + E_\Sigma \left[ \mathbf{\Sigma}_x^2 - \mathbf{\Sigma}_y^2 \right] + \mathbf{\Sigma} \cdot \mathbf{A} \cdot (\mathbf{I_1} + \mathbf{I_2}), \qquad (4a)$$

$$D_\Sigma = 3\alpha_\Sigma D_e + \beta_\Sigma D_c, \quad E_\Sigma = \alpha_\Sigma E_e + \beta_\Sigma E_c, \quad \alpha_1 = {17}/{10}, \quad \beta_1 = -{12}/{5} \qquad (4b)$$

where $\Sigma$ represent spin operators for the exchange-coupled $Mn^{4+}$ ions, the first term describe the electron Zeeman interaction, the second and third ones describes the axial and orthorhombic second rank fine splitting, $A$ are the magnetic hyperfine interaction constants of the 100% abundant isotope $^{55}Mn$, $I_1=I_2=5/2$. The main axes $z$, $x$ and $y$ of second rank fine structure tensor are parallel to $[110]$, $[\bar{1}10]$ and $[001]$ directions respectively. If the $Mn^{4+}$ is located along a [110] direction relative to the other substitutional $Mn^{4+}$ ion than it generates an orthorhombic field and is related to six other symmetry conjugated sites. The spin Hamiltonian parameters for the $\Sigma = 1$ transitions presented in Table 3 were derived from angular dependences of EPR spectra recorded with the magnetic field varied in (100) and (110) planes. As shown on Fig. 6 the angular dependence of $\Sigma = 1$ is strongly defined by the both $D_1$ and $E_1$ terms. We have found that $E_1 / D_1 \approx 1/3$ that means there is maximum rhombic splitting and the components of corresponding Cartesian second rank fine tensor $\mathbf{D}$ are the following [26]:

$$D_{xx} = -{D_1}/{3} + E_1 \approx 0, \qquad D_{yy} = -{D_1}/{3} - E_1 \approx -{2D_1}/{3}, \qquad D_{zz} = {2D_1}/{3} \qquad (5)$$

For this dimer centers which exhibits isotropic g-factor the anisotropy of exchange interaction is mainly caused by dipole-dipole interaction. If we adopt the point dipole approximation for the $Mn^{4+}$ ions constituting the pair then the value of $D_e$ is defined as follows: $D_e = -{g^2\beta^2}/{R^3}$, where R is interionic distance. By use of formulas (4b) and suggesting that axial splitting of isolated $Mn^{4+}$ ion $D_c \sim 0$ we obtain the room temperature value of $D_e$~-0.0147 cm$^{-1}$ and corresponding distance of approximately 4.9 Å. This value is significantly reduced from the separation of ~5.5 Å between nearest regular Ti$^{4+}$ sites located along <110> direction. These findings are confirmed experimentally by the observation of characteristic hyperfine structure of the exchange-coupled pair complexes as shown on Fig.7. The hyperfine structure consist of 11 lines with hyperfine splitting constant half that of the isolated Mn$^{4+}$ ions and with intensity ratio 1:2:3:4:5:6:5:4:3:2:1 expected for identical Mn$^{4+}$ ions coupled by an isotropic exchange interaction larger than the hyperfine interaction. The intensities of the $\Sigma = 1$ transitions are difficult to follow as a function of temperature and therefore the energy intervals between the spin states $\Sigma$ were not evaluated.

**Table 3.** The Spin Hamiltonian parameters of single Mn$^{4+}$ centres and Mn$^{4+}$- Mn$^{4+}$ ($\Sigma = 1$) exchange-coupled pairs in SrTiO$_3$ single crystals (in 10$^{-4}$ cm$^{-1}$), T=295K.

| centre | $D_{\Sigma=1}$ | $E_{\Sigma=1}$ | g | A | ref. |
|---|---|---|---|---|---|
| Mn$^{4+}$ | - | - | 1.994 | -69.4 | [24] |
| Mn$^{4+}$- Mn$^{4+}$ | -750.0 | -248.0 | 1.994 | -34.7 | * |

* present study

The analysis of the angular variation of EPR spectra of Mn$^{4+}$ dimers provides a basis for constructing the model of exchange coupled pairs in SrTiO$_3$ lattice. The structure of next nearest neighbor pair site, shown on Fig. 8, has point group symmetry D$_{2h}$ in a cubic phase of SrTiO$_3$. The pairs consist of two Mn$^{4+}$ ions which are substituted for nearest Ti$^{4+}$ sites located along one of the <110> crystal directions. These measurements do not provide complete information related to the energy level sequence of Mn$^{4+}$- Mn$^{4+}$ pairs but the data obtained do suggest that the $\Sigma = 1$ state is higher in energy in comparison to the other $\Sigma = 2, 3$ states. The magnetic interactions must be transmitted through pathways involving three intervening ions: one titanium and two oxygen atoms. In this case ferromagnetic right angle superexchange interaction is expected because of Mn-O bonds in perovskite structure.

## 4. Conclusions

The detailed studies have performed on the several complexes of high spin Mn$^{3+}$ ions. We were able to define the sign and value of the axial fine structure parameters $b_2^0$ that is important to elucidate the defect structure at the $Mn^{3+}$ site. The tetragonally elongated octahedron coming from Jahn-Teller distortions of a local site of $Mn^{3+}$ centers have been ascertained from EPR data. Spin density at the nucleus for manganese in the distorted octahedral coordination has been found considerably different for various types of the centers. A variety of complex $Mn^{3+}$ ions detected in $SrTiO_3$ crystals has been ascribed to differences in charge compensation. These variations are largely due to the techniques used to grow and dope the $SrTiO_3$ single crystals. Our results show the formation of exchange coupled $Mn^{4+}$ dimers substituted for the second nearest neighbor $Ti^{4+}$ sites in $SrTiO_3$ crystals.

## Acknowledgements

This work was supported by the Operational Program Research and Development for Innovations - European Social Fund (project CZ.1.05/2.1.00/03.0058 of the Ministry of Education, Youth and Sports of the Czech Republic), project SAFMAT CZ.2.16/3.1.00/22132, grant 202/09/J017 of GACR and grant No. 01010517 of the TACR, by the PP RAS "Quantum physics of condensed Matter" and by the RFBR No. 12-02-00848, by the Ministry of Education and Science under Contract No. 14.740.11.0048, by the programs of RAS: "Spintronics".

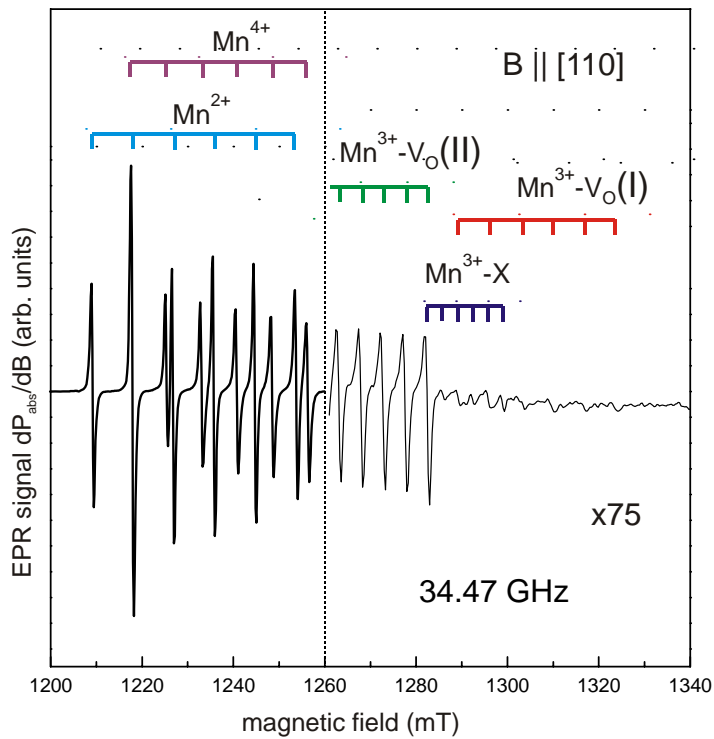

**Figure 1.** Q-band EPR spectra of SrTiO$_3$:Mn showing the cubic Mn$^{4+}$ and Mn$^{2+}$ as well as charge compensated Mn$^{3+}$ ions. B|| [100], $T$ = 295K. EPR spectrum of axial Mn$^{3+}$ centres is enlarged by x75 in comparison to one of the cubic Mn$^{4+}$ and Mn$^{2+}$. The resolved hyperfine structure shown by the cropped lines is due to the interaction with the $^{55}$Mn nuclei.

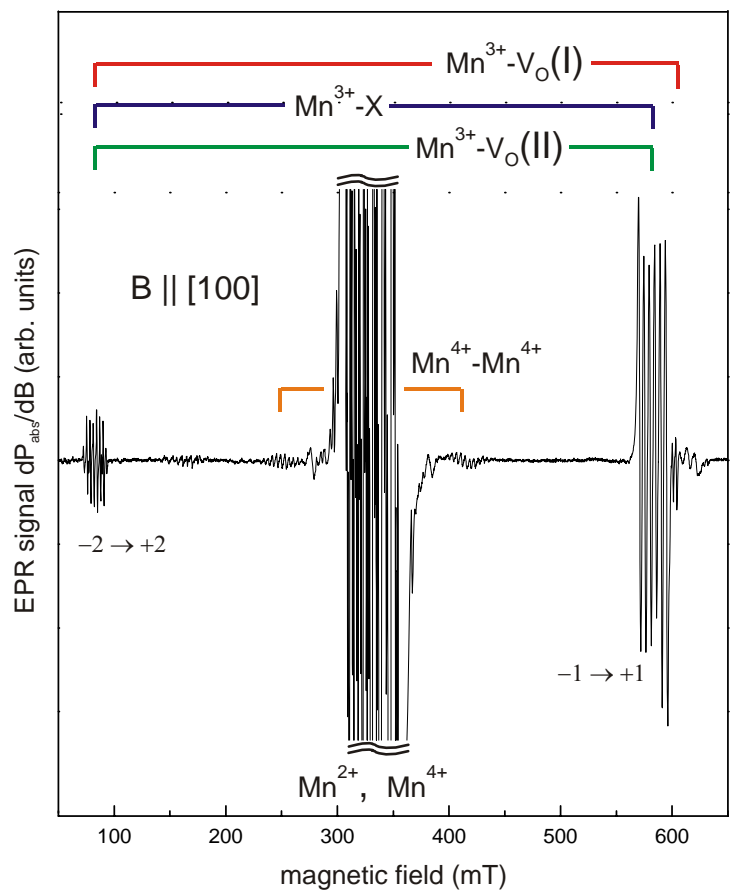

**Figure 2.** X-band EPR spectrum of the charge compensated $Mn^{3+}$ ions and $Mn^{4+}$ dimers when applied magnetic field parallel to [100] direction, microwave frequency of 9.24 GHz, $T = 295K$. EPR spectra of cubic $Mn^{2+}$ and $Mn^{4+}$ ions are cut because of their extremely strong intensity. The fine structure transitions are shown by the cropped lines.

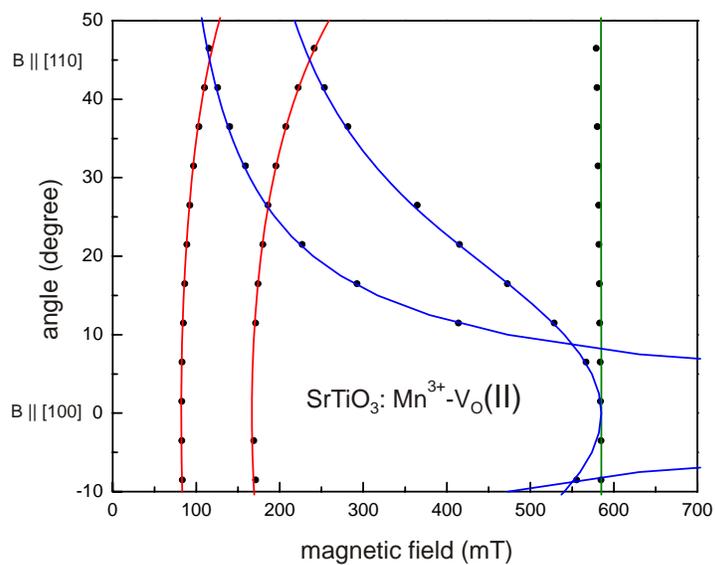

**Figure 3.** The EPR road map of the center of gravity of $Mn^{3+}$-$V_O$(II) resonance lines when applied magnetic field varies in the (100) plane, microwave frequency of 9.26 GHz, T = 150 K. The lines are the fine structure transitions calculated by exact diagonalization of spin Hamiltonian (1). The cycles show the experimental points.

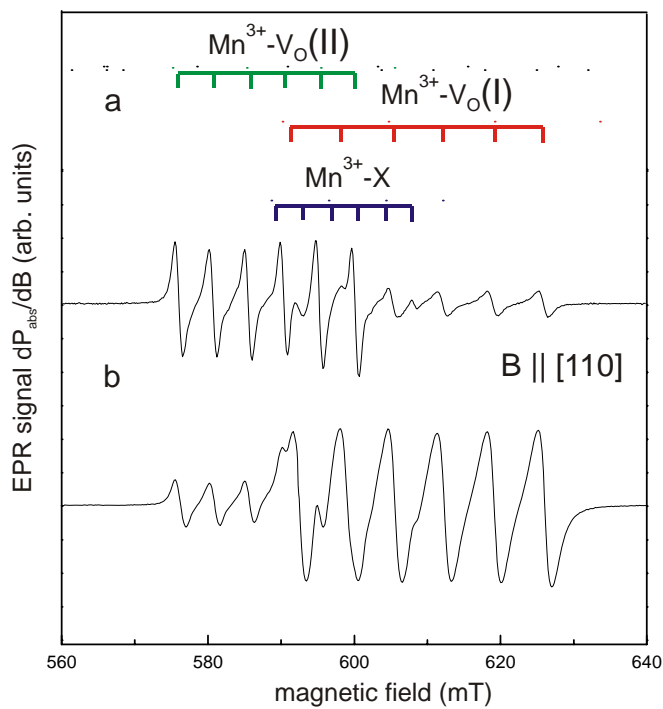

**Figure 4.** EPR spectrum of as grown (a) and additionally reduced (b) manganese-doped $SrTiO_3$ with the applied magnetic field parallel to the [110] axis, microwave frequency of 9.25 GHz, $T$ = 295 K. The spectrum consists of lines corresponding to the 1 ↔ -1 transition within $^5B_1$ ground state of distinct types of $Mn^{3+}$ centers. The six fold splitting induced by the hyperfine interaction with the Mn nuclear spin $I$ = 5/2 is shown by the cropped lines. The increase of intensity of $Mn^{3+}$-$V_O$(I) resonance lines in comparison to $Mn^{3+}$-$V_O$(II) once became apparent from these spectra.

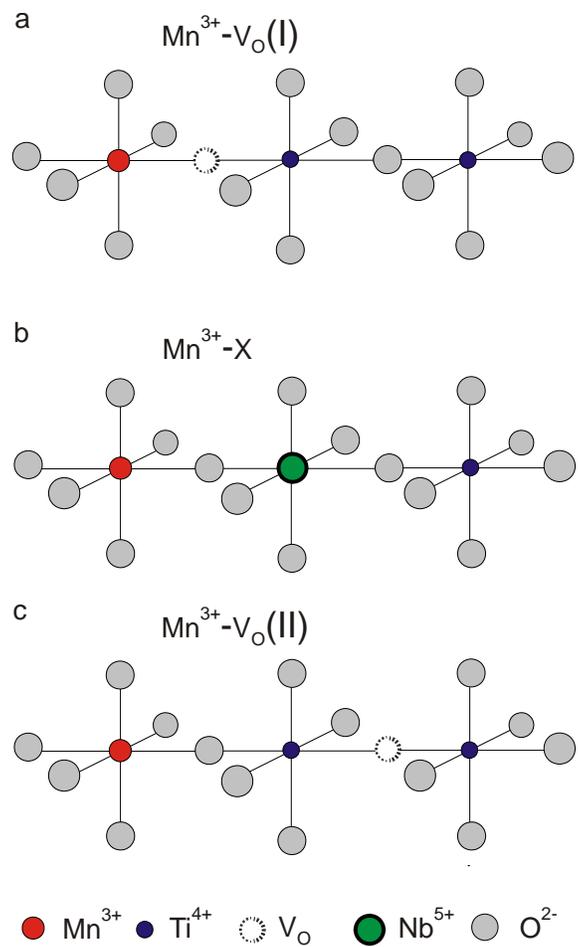

**Figure 5.** Possible models of charge compensated (a) $Mn^{3+}$-$V_O$(I), (b) $Mn^{3+}$-X and (c) $Mn^{3+}$-$V_O$(II) centers in $SrTiO_3$ structure including the probable presence of a charge compensator along a [100] direction in the first, second ($Nb^{5+}_{Ti}$) and third coordination sphere.

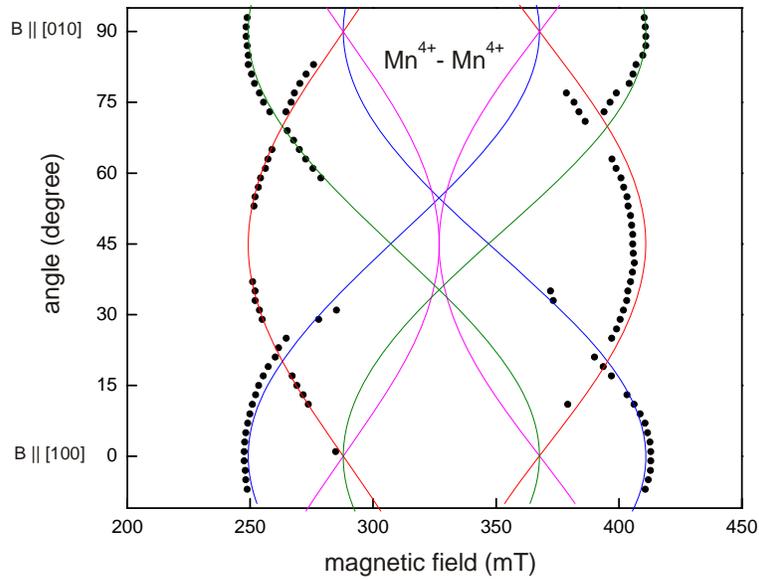

**Figure 6.** The road map of the center of gravity of EPR lines of $\Sigma = 1$ transitions from the $Mn^{4+}$- $Mn^{4+}$ exchange-coupled pairs when applied magnetic field varies in the (100) plane, $T = 295$ K. The lines are the calculated positions of the resonances.

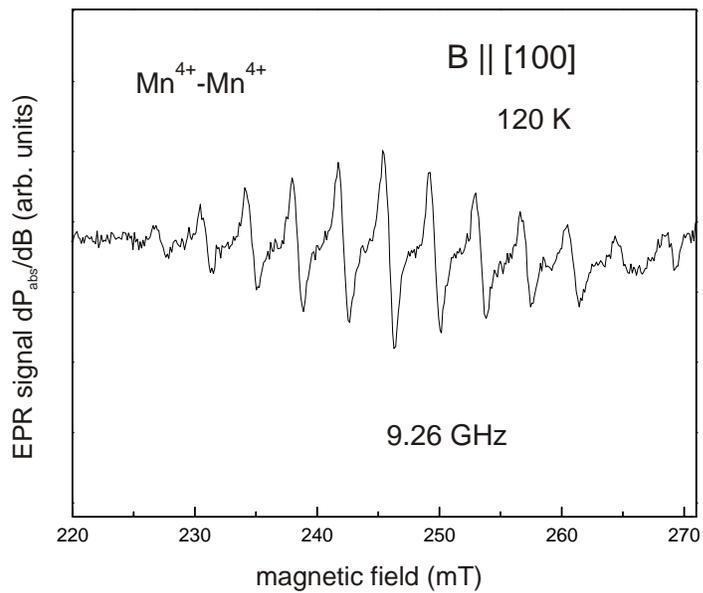

**Figure 7.** Hyperfine structure in the EPR spectrum of the next nearest neighbor exchange-coupled $Mn^{4+}$ - $Mn^{4+}$ pairs. This structure could be uniquely assigned to two equivalent $Mn^{4+}$ ions ($I=5/2$).

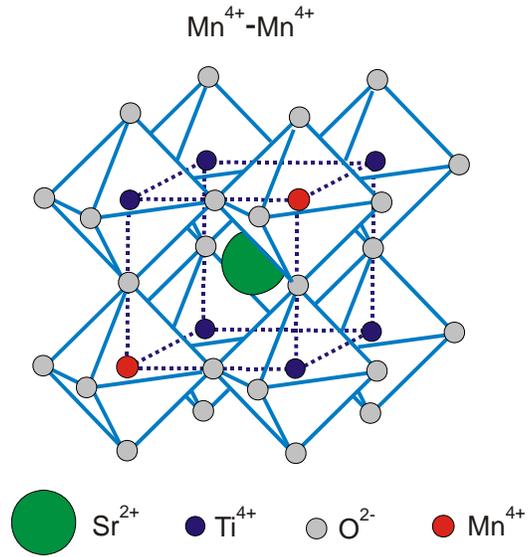

**Figure 8.** Model of exchange-coupled $Mn^{4+}$ pairs in the cubic phase of $SrTiO_3$ with two $Mn^{4+}$ ions which are substituted for nearest $Ti^{4+}$ sites located along one of the <110> directions. The magnetic dipolar interaction between $Mn^{4+}$ ions gives rise to anisotropy of exchange interaction.